\documentclass[11pt]{article}
\usepackage{amssymb}
\usepackage{amsfonts}
\usepackage{amsmath}

\begin{document}

\title{\textbf{Renormalization-group improved inflationary scenarios}}
 \author{E.O.~Pozdeeva\footnote{E-mail: pozdeeva@www-hep.sinp.msu.ru}, S.Yu.~Vernov\footnote{E-mail: svernov@theory.sinp.msu.ru}\\
Skobeltsyn Institute of Nuclear Physics, Lomonosov Moscow State University,\\ Leninskie Gory~1, 119991, Moscow, Russia}
\date{ \ }
\maketitle

\begin{abstract}
The possibility to construct an inflationary scenario for
renormalization-group improved
potentials corresponding to the Higgs sector of quantum field models is
investigated. Taking into account quantum corrections to the
renormalization-group potential which sums all leading logs of
perturbation theory is essential
for a successful realization of the inflationary scenario, with very reasonable
parameters values. The scalar electrodynamics inflationary scenario thus obtained are seen to be in good agreement with
the most recent observational data.
\end{abstract}


\section{Introduction}

The evolution of the Universe can be separated on four main epochs: inflation, radiation domination stage, matter domination stage and dark energy domination stage. The existence of stage of accelerated expansion in the early Universe
(inflation) provides a simple explanation of astronomical data, including the fact that at very large distances the Universe is approximately isotropic, homogeneous and spatially flat. The resent observations result in important restrictions on  inflationary models. In particular, the high-precision measurements by the Planck telescope~\cite{Planck2015} show that the single-field inflationary models are realistic, but simplest inflationary models with minimally coupled scalar fields should be ruled out.

 Adding a tiny non-minimal coupling of the inflaton field to gravity can essentially change the observation predictions of the inflationary model. In such a way it is possible to construct inflationary scenarios compatible with observation data~\cite{nonmin-infl,HiggsInflation,DeSimone:2008ei,betainfl}. The resent observation data support to take into account quantum properties of the inflaton, because quantum corrections to the action of the scalar field minimally coupled to gravity include non-minimally coupling term~\cite{ChernikovTagirov} and   it is necessary to introduce an induced gravity term proportional to $R\sigma^2$, where $\sigma$ is a scalar field, in order to renormalize the quantum scalar field theory in curved space-time~\cite{Callan:1970ze}.

   Quantum field theory in the curved  space-time is an important concept at the very early universe, where curvature is large and typical energies are very high.
 (for a complete description, see~\cite{BOS}). As a consequence, it is natural to start by addressing the issue of the
renormalization-group (RG)  improved effective potential for an arbitrary renormalizable
massless gauge theory in curved space-time~\cite{Elizalde:1993ee,Elizalde:1994im}.

In~\cite{EOPV2014,EOPV2015} inflationary models with RG-improved effective potentials  proposed in~\cite{Elizalde:1993ee,Elizalde:1994im} have been constructed.
 It has been found that inflation can be realized for scalar electrodynamics, $SU(5)$ gauge models and wide class of finite gauge models due to the use of RG-improved potentials~\cite{EOPV2014,EOPV2015}. Inflationary model based on the $\sigma^4$ RG-improved potential has been considered in~\cite{Inagaki:2014wva}.
 In the paper~\cite{EOPV2015} it has been also shown that RG-corrections can  generate the Hilbert--Einstein term in the action. In other words, we do not need to include this term by hand, as is done in Higgs-driven inflation~\cite{HiggsInflation,DeSimone:2008ei}, since we get it naturally, as part of the quantum corrections to the induced gravity term. One of the motivations for our research is to show that using quantum corrections it is possible to construct inflationary scenarios  without adding  the Hilbert--Einstein term in the action. In this paper we describe main steps of the inflationary scenario construction on the example of the scalar electrodynamics. We obtain useful analytic expressions for the inflationary parameters and the number of e-foldings during inflation.

\section{Inflationary models with non-minimal coupling and RG-improved potentials}

To construct inflation scenario we consider the following action:
\begin{equation}
\label{action} S=\int d^4 x \sqrt{-g}\left[
U(\sigma)R-\frac12g^{\mu\nu}\sigma_{,\mu}\sigma_{,\nu}-V(\sigma)\right],
\end{equation}
where $U(\sigma)$ and $V(\sigma)$ are differentiable functions.

For a spatially flat FLRW  metric with the  interval
$ds^2=-dt^2+a^2(t)\!\left[dx_1^2+dx_2^2+dx_3^2\right]\!$,
the Einstein equations have the following form~\cite{KTVV2012}:
\begin{equation}
\label{Fr1} 6UH^2+6\dot U H=\frac{1}{2}\dot\sigma^2+V\,,\quad
2U\left(2\dot H+3H^2\right)+4\dot U H+2\ddot U={}-\frac{1}{2}\dot\sigma^2+V\,,
\end{equation}
where the Hubble parameter is the logarithmic derivative of the
scale factor: $H=\dot a/a$ and differentiation with respect to
time $t$ is denoted by an overdot.

In the case of quasi de Sitter expansion there is no difference between
spectral indexes calculated either in the Jordan frame directly, or in the Einstein frame after conformal transformation~\cite{Kaiser}. By this reason
we calculate inflationary parameters using the corresponding model in the Einstein frame,
obtained from model (\ref{action})  by conformal transformation of the metric
$\tilde{g}_{\mu\nu} = 16\pi M_{\mathrm{Pl}}^{-2} U(\sigma) g_{\mu\nu}$. The metric in the Einstein frame is marked with a tilde, $M_{\mathrm{Pl}}$ is the Planck mass.
After this transformation, we get a model for a minimally coupled scalar field,
described by the following action
\begin{equation}
S_E =\int d^4x\sqrt{-\tilde{g}}\left[\frac{M_{\mathrm{Pl}}^2}{16\pi}R(\tilde{g}) -
\frac{Q(\sigma)}{2}\tilde{g}^{\mu\nu}\sigma_{,\mu}\sigma_{,\nu}-
V_\mathrm{E}(\sigma)\right],
\label{action1}
\end{equation}
where
\begin{equation}
Q=\frac{M_{\mathrm{Pl}}^2\left(U+3U'^2\right)}{16\pi U^2},\qquad  V_\mathrm{E} = \frac{
M_{\mathrm{Pl}}^4V}{256\pi^2U^2}.
\label{QVE}
\end{equation}

Many inflationary models are based upon the possibility of a slow evolution of scalar
field~$\sigma$ in the neighborhood of the unstable de Sitter solution of~(\ref{Fr1}). In any frame the value of the scalar field at a de Sitter solution $\sigma_f$ is defined by the condition $V'_\mathrm{E}(\sigma_f)=0$. If $U(\sigma_f)>0$ and $V''_\mathrm{E}(\sigma_f)<0$, then  the corresponding de Sitter solution is unstable~\cite{Skugoreva:2014}.

In this paper  we consider  the cosmological model defined by the scalar electrodynamics  RG-improved potential~\cite{Elizalde:1993ee,EOPV2015}:
\begin{equation}
\label{VUSCED}
V=\frac{\lambda\sigma^4}{4!}+\frac{6e^4\sigma^4}{(8\pi)^2}\vartheta,
\quad
U=\frac{\xi\sigma^2}{2}+\frac{2e^2\sigma^2}{(8\pi)^2}\vartheta, \quad \mbox{where} \quad\vartheta=\frac{1}{2} \ln \left(\frac{\sigma^2}{\mu^2}\right).
\end{equation}
$\lambda$, $\xi$,  $e$ and $\mu$ are constants. The second summands of $V$ and $U$, proportional to $\vartheta$, are RG-corrections. It is easy to see that without these corrections  one gets a constant potential $V_\mathrm{E}$ that is not suitable for inflation.
It has been shown in~\cite{EOPV2014} that at $\lambda=18 e^2\xi$ one obtains unstable de Sitter solutions with $\sigma_f=\pm\mu$, therefore, proportional to  $\vartheta(\sigma)$ RG-corrections are equal to zero at the de Sitter point.

It is interesting that the function $\vartheta(\sigma)$ can be used as a scalar field during inflation. Indeed, we get the potential
\begin{equation}\label{VESCED}
V_\mathrm{E}=\frac{3M_{\mathrm{Pl}}^4e^2(8\pi^2\xi+e^2\vartheta)}{8(16\pi^2\xi+e^2\vartheta)^2}.
\end{equation}
During inflation parameters of the slow-roll approximation~\cite{DeSimone:2008ei,Kaiser,Liddle}, that should remain to be less than one, can be expressed as follows:
\begin{equation}
\label{SLP_phi}
\begin{split}
\epsilon &= \frac{M_{\mathrm{Pl}}^2{(V_\mathrm{E}')}^2}{16\pi V_\mathrm{E}^2Q}=\frac{{e}^{8}
{\vartheta}^{2}}{\left(8\pi^2\xi+e^2\vartheta\right)^2(12e^4\vartheta^2+4{e}^{2}\beta_2\vartheta+ \beta_0)},\\
\eta &=\frac{M_{\mathrm{Pl}}^2}{8\pi V_\mathrm{E}Q}\left[V_\mathrm{E}''-\frac{V_\mathrm{E}'Q'}{2Q}\right]=\frac {2{e}^{4} \left( 24{e}^{6}{\vartheta}^{3}+6{e}^{4} \beta_2 {\vartheta}^{2}+{e}^{2} \beta_1 \vartheta-16{\pi}^{2}\xi \beta_0
 \right) }{ \left( 8{\pi }^{2}\xi+{e}^{2}\vartheta \right)
 \left( 12{e}^{4}{\vartheta}^{2}+4{e}^{2} \beta_1 \vartheta+\beta_0 \right)^2},
\end{split}
\end{equation}
where we introduce the following constants:  $ \  \beta_0 =3072\pi^4\xi^2+512\pi^4\xi+192\pi^2e^2\xi+3e^4$,
\begin{equation*}
\beta_1=256\pi^4\xi+96\pi^{2}e^2\xi+3e^4,\qquad
\beta_2=96\pi^{2}\xi+8\pi^{2}+3e^2\,.
\end{equation*}

Therefore, the most important inflationary parameters, namely, the ratio  of squared amplitudes of tensor and scalar perturbations $r= 16 \epsilon$ and the scalar spectral index of the primordial curvature fluctuations~$n_\mathrm{s}= 1 - 6 \epsilon + 2\eta$, are functions of $\vartheta$. Moreover, the number of e-foldings during inflation  is a function of $\vartheta$ as well:
\begin{equation*}
\begin{split}
N_e&=
\frac{8\pi}{M_{\mathrm{Pl}}^2}\!\int\limits_{\sigma_{\mathrm{end}}}^{\sigma_{\mathrm{i}}}
\left|\frac{V_\mathrm{E}}{V'_\mathrm{E}}\right|Q\,d\tilde\sigma=\!\int\limits_{\vartheta_{\mathrm{end}}}^{\vartheta_{\mathrm{i}}}\!\left|
\frac{\sqrt{12e^4\tilde{\vartheta}^2+4e^2\beta_2\tilde{\vartheta}+\beta_0}}{(32\pi^2\xi+2e^2\tilde{\vartheta})\sqrt{\epsilon}}\right|\,d\tilde\vartheta=\\
&=\left|\left.\frac {3e^4\ln  \left( 32\,{\pi }^{2}\xi+2\,{e}^{2}\vartheta \right)
+ c_0 \ln  \left( 2\vartheta \right) +2e^{2}\left( 2{e}^{2}\vartheta+4{e}^{2}+96\,{\pi }^{2}
 \left( \xi+1/9 \right)  \right) \vartheta }{4{e}^{4}}\right|_{\vartheta_{\mathrm{end}}}^{\vartheta_{\mathrm{i}}}\right|
\end{split}
\end{equation*}
where
$c_0=\left( {e}^{4}+64\,{\pi }^{2}\xi\,{e}^{2}+1024\,{\pi }^{4}\xi
\, \left( \xi+1/6 \right)  \right)$,
 $\sigma_{\mathrm{end}}$ is the value of the field at the end of inflation, defined by the condition $\epsilon=1$ at $\sigma=\sigma_{\mathrm{end}}$. The value of $\sigma_{\mathrm{i}}$ defines the moment at which the inflationary parameters are calculated. The suitable values of $N_e$ is from $50$ to $65$.

The obtained analytic formulae essentially simplify the search of the model parameters suitable for realistic inflationary scenario. Some examples have been found in~\cite{EOPV2014} due to numerical calculations. In this paper we generalize this result and obtain new suitable values for parameters $\xi$ and $e$ (see Table~\ref{ScalarElectroDinPar}).
\begin{table}[h]
\begin{center}
\caption{Parameter values for the scalar electrodynamics inflationary scenario.}
\begin{tabular}{|c|c|c|c|c|c|c|c|}
  \hline
   $\xi$& $e$&  $\vartheta_{end}$ ($\epsilon=1$)& $N_e$ & $\vartheta_N$ & $n_\mathrm{s}$ & $r$ \\
   \hline
  $1$ & $5$ & $-2.9592$ & $50$ & $-1.11735$ & $0.9600$ & $0.0105$  \\
  $1$ & $5$ & $-2.9592$ & $55$ & $-1.05767$ & $0.9628$ & $0.0087$  \\
  $1$ & $5$ & $-2.9592$ & $60$ & $-1.00306$ & $0.9651$ & $0.0073$  \\
  $1$ & $5$ & $-2.9592$ & $65$ & $-0.95284$ & $0.9671$ & $0.0062$ \\
  \hline
  $5$ & $10$ & $-3.7234$ & $50$ & $-1.60738$ & $0.9623$ & $0.0132$  \\
  $5$ & $10$ & $-3.7234$ & $55$ & $-1.53516$ & $0.9651$ & $0.0111$ \\
  $5$ & $10$ & $-3.7234$ & $60$ & $-1.46869$ & $0.9675$ & $0.0094$ \\
  $5$ & $10$ & $-3.7234$ & $65$ & $-1.40720$ & $0.9695$ & $0.0081$  \\
  \hline
  $10$ & $10$ & $-7.6441$ & $50$ & $-4.9426$ & $0.9676$ & $0.0284$  \\
  $10$ & $10$ & $-7.6441$ & $55$ & $-4.8316$ & $0.9703$ & $0.0247$ \\
  $10$ & $10$ & $-7.6441$ & $60$ & $-4.7276$ & $0.9726$ & $0.0217$ \\
  $10$ & $10$ & $-7.6441$ & $65$ & $-4.6299$ & $0.9745$ & $0.0193$  \\
  \hline
  $10$ & $15$ & $-3.2876$ & $50$ & $-1.26908$ & $0.9606$ & $0.0108$  \\
  $10$ & $15$ & $-3.2876$ & $55$ & $-1.20380$ & $0.9634$ & $0.0090$ \\
  $10$ & $15$ & $-3.2876$ & $60$ & $-1.14402$ & $0.9658$ & $0.0075$ \\
  $10$ & $15$ & $-3.2876$ & $65$ & $-1.08897$ & $0.9678$ & $0.0064$  \\
  \hline
\end{tabular}
\label{ScalarElectroDinPar}
\end{center}
\end{table}

\section{Conclusions}
In this paper we continue to consider the inflationary models with non-minimal coupling and renormalization-group improved potentials. In our previous work~\cite{EOPV2014} we construct inflationary model for the scalar electrodynamics using numerical calculations of the inflationary parameters. In this paper we show that both slow-roll parameters and the number of e-foldings during inflation are elementary functions of $\vartheta$ that defines RG-corrections. The obtained analytic formulae allows us to enlarge the domain of model parameters that is suitable for inflation. It is interesting that the RG equation and $\beta$-functions also arise as part of a the method of the inflationary scenario construction~\cite{betainfl}.

We would like to thank Emilio Elizalde and Sergei Odintsov  for collaboration and fruitful discussions.  Research of E.O.P. is supported in part by grant MK-7835.2016.2  of the President of Russian Federation. Research of S.Yu.V. are supported in part by RFBR grant 14-01-00707.

\end{document}